\documentclass[prd,showpacs,superscriptaddress,twocolumn,
floatfix,preprintnumbers,altaffilletter]{revtex4}

\usepackage{longtable}
\usepackage{graphicx}
\usepackage{amsmath,amssymb}
\usepackage{color}
\usepackage{units}
\usepackage{epstopdf}
\usepackage{hyperref}

\newcommand{\beq}{\begin{equation}}
\newcommand{\eeq}{\end{equation}}
\newcommand{\bdm}{\begin{displaymath}}
\newcommand{\edm}{\end{displaymath}}

\definecolor{Gray}{gray}{0.9}

\graphicspath{{./plots/}}
\begin{document}


\title{The detectability of eccentric compact binary coalescences with advanced gravitational-wave detectors}
\author{M Coughlin}
\email{coughlin@physics.harvard.edu}
\affiliation{Department of Physics, Harvard University, Cambridge, MA 02138, USA}
\author{P Meyers}
\email{meyers@physics.umn.edu}
\affiliation{School of Physics and Astronomy, University of Minnesota, Minneapolis, Minnesota 55455, USA}
\author{E Thrane}
\affiliation{LIGO Laboratory, California Institute of Technology, MS 100-36,
Pasadena, CA, 91125, USA}
\affiliation{School of Physics, Monash University, Clayton, Victoria 3800, Australia}
\author{J Luo}
\affiliation{Physics and Astronomy, Carleton College, Northfield, Minnesota 55057}
\author{N Christensen}
\affiliation{Physics and Astronomy, Carleton College, Northfield, Minnesota 55057}
\date{\today}

\begin{abstract}
Compact binary coalescences are a promising source of gravitational waves for second-generation interferometric gravitational-wave detectors such as advanced LIGO and advanced Virgo.
While most binaries are expected to possess circular orbits, some may be eccentric, for example, if they are formed through dynamical capture.
Eccentric orbits can create difficulty for matched filtering searches due to the challenges of creating effective template banks to detect these signals.
In previous work, we showed how seedless clustering can be used to detect low-mass ($M_\text{total}\leq10M_\odot$) compact binary coalescences for both spinning and eccentric systems, assuming a circular post-Newtonian expansion.
Here, we describe a parameterization that is designed to maximize sensitivity to low-eccentricity ($0\leq\epsilon\leq0.6$) systems, derived from the analytic equations.
We show that this parameterization provides a robust and computationally efficient method for detecting eccentric low-mass compact binaries.
Based on these results, we conclude that advanced detectors will have a chance of detecting eccentric binaries if optimistic models prove true.
However, a null observation is unlikely to firmly rule out models of eccentric binary populations.
\end{abstract}

\maketitle

\section{Introduction}\label{sec:Intro}
Compact binary coalescences (CBCs) of black holes and/or neutron stars are a likely source of gravitational waves (GWs) \cite{S6Highmass,S6Lowmass,0264-9381-27-17-173001}.
The GWs generated by CBCs, such as binary neutron stars (BNSs), neutron-star black holes (NSBHs), and binary black holes (BBHs), sweep upward in frequency and strain amplitude through the sensitive band of GW detectors such as LIGO~\cite{aLIGO}, Virgo~\cite{VIRGO}, GEO~\cite{GEO600}, and KAGRA~\cite{kagra}.
Their detection will provide information about the populations of compact objects in the universe \cite{0004-637X-676-2-1162}, provide a probe for the properties of strong field gravity, and are a way to test general relativity \cite{PhysRevD.89.082001}.

Searches for CBCs almost entirely use matched filtering, which requires precise knowledge of astrophysical waveforms. Excess power searches are also used, predominantly for high-mass systems that result in shorter signals~\cite{imbh,0264-9381-25-11-114029}.
Matched filtering provides a nearly optimal strategy for detecting compact binaries because they are well-modeled systems. Due to computational limitations, most CBC searches use template banks composed of non-spinning, non-eccentric waveforms, which are less computationally challenging to implement than searches with complications such as spin and eccentricity. Up until now, there have been no dedicated matched filtering searches for low mass binaries with low to moderate eccentricities. Huerta and Brown have shown that searches using waveforms that assume no eccentricity are significantly sub-optimal above $\epsilon > 0.05$ \cite{PhysRevD.87.127501}.
They conclude that in order to detect and study the rate of eccentric stellar-mass compact binaries in aLIGO, a search specifically targeting these systems will need to be constructed. Matched filtering searches would require eccentric template banks to avoid being significantly sub-optimal.
One method for overcoming these difficulties is building larger (and smarter) template banks to perform searches. This is, of course, computationally expensive, and in some cases, intractable.

There have been a number of recent developments that potentially enable searches for eccentric binaries.
Huerta et al.~\cite{HuKu2014} recently developed a purely analytic, frequency-domain model for gravitational waves emitted by compact binaries on orbits with small eccentricity. This model reduces to the quasi-circular post-Newtonian approximant TaylorF2 at zero eccentricity and to the post-circular approximation of Yunes et al.~\cite{PhysRevD.80.084001} at small eccentricity. 
A computationally-tractable, matched filtering search using these templates is possible.
Matched filtering searches rely on knowing the phase of the gravitational-wave signal being searched for. This is powerful for limiting the noise background of detector data but also is subject to modeling errors, especially in highly eccentric cases where perturbative waveform generation methods are not yet sufficiently accurate to be used as templates.
Another proposed method for detecting highly eccentric binaries is the search for the ``repeated bursts'' created by the many passes of the eccentric binary \cite{PhysRevD.85.123005,PhysRevD.90.103001}. 
Tai et al. \cite{PhysRevD.90.103001} applied a single-detector power stacking algorithm, developed in \cite{KaCa2009} to search for gravitational-wave bursts associated with soft gamma ray repeater events, to the case of eccentric binary mergers. They use a time-frequency signature informed by an eccentric model developed in \cite{EaMc2013} to sum up power in Q-transform pixels, which is a multi-resolution basis of windowed complex exponentials.
Excess power methods, like those from \cite{PhysRevD.90.103001} and those presented below, do not have the same issues as matched filtering, as phase information is not used in these analyses.
In the work that follows, we will differ from \cite{PhysRevD.90.103001} in the use of a coherent multi-detector statistic with a generic eccentric frequency-time track with the assumption of low-to-moderate eccentricity CBCs.

There is significant astrophysical motivation for designing searches for eccentric CBCs. Main sequence evolution binaries will be circularized by the time they enter the sensitive frequency band for ground-based detectors \cite{PhysRevD.81.024007}. On the other hand, models exist which could result in highly eccentric CBCs in the sensitive band. O'Leary et al.~present a model where CBCs with high eccentricities can be produced by the scattering of stellar mass black holes in galactic cores containing a super-massive black hole \cite{ScatteringBlackHoles}. They show that 90\% of such systems would have eccentricity $\epsilon > 0.9$ when entering the sensitive band. 
Assuming a highly idealized pipeline, and ignoring complications such as template bank trial factors, they found the expected rate of coalescence detectable by aLIGO to be $~ 1-10^2$ per year.
In this work, we argue that the detection rate in realistic pipelines is probably closer to $0.001$--$0.5$ per year.

Samsing et al. \cite{0004-637X-784-1-71} show how interactions between compact binaries and single objects can induce chaotic resonances in the binary system and create eccentric binaries. Although the number of BBHs in the galactic center is not well constrained, there may be more than $10^3$\,black holes in central 0.1\,pc of our galaxy \cite{1538-4357-645-2-L133}. Binary-binary interactions in globular clusters can also result in non-zero eccentricity. If the orbital planes of the inner and outer binary are highly inclined with respect to one another, Kozai resonances increase the eccentricity of the inner binary.
It is estimated that 30\% of binaries formed in this way will have eccentricities $\epsilon > 0.1$ when they enter the sensitive band~\cite{PhysRevD.81.024007}. Eccentric binary black hole mergers have also been studied \cite{PhysRevD.78.064069,escidoc:795602}. In the presence of an accretion disk, a black hole orbitting counter to the motion of the disk maintains its eccentricity (rotating with the disk has the opposite effect). Numerical relativity simulations have shown that highly eccentric BNS systems can exhibit interesting features, including f-modes and disks resulting from the merger \cite{Gold:2011df}. Simulations of NSBH mergers show varying amounts of mass transfer and accretion disk size \cite{East:2011xa,Stephens:2011as}. Therefore, there is significant motivation to design searches for eccentric CBCs as GW sources. Another potential mechanism for forming eccentric neutron star binaries is tidal capture \cite{0004-637X-720-1-953}.

\begin{figure*}[t]
 \includegraphics[width=3.5in]{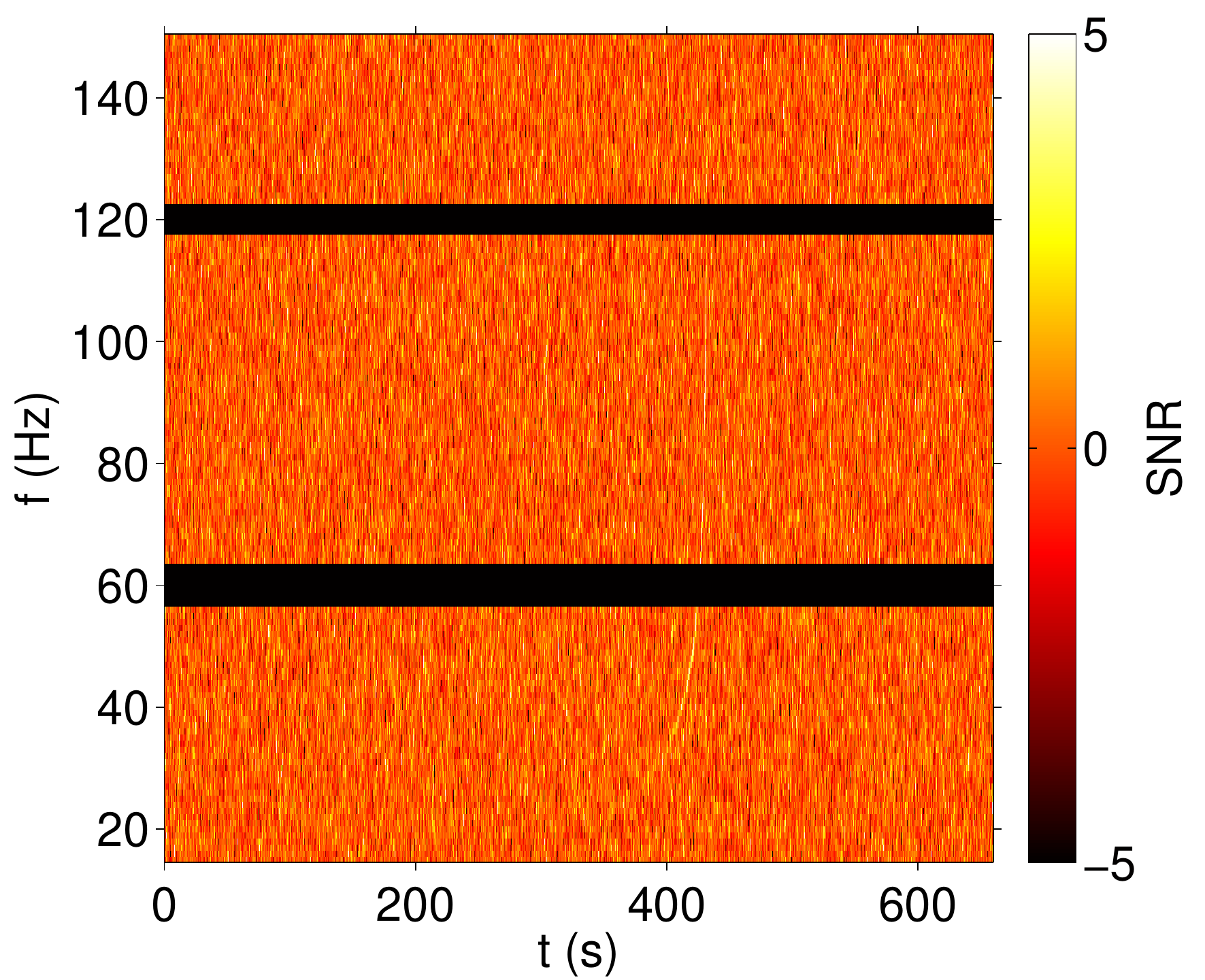}
 \includegraphics[width=3.5in]{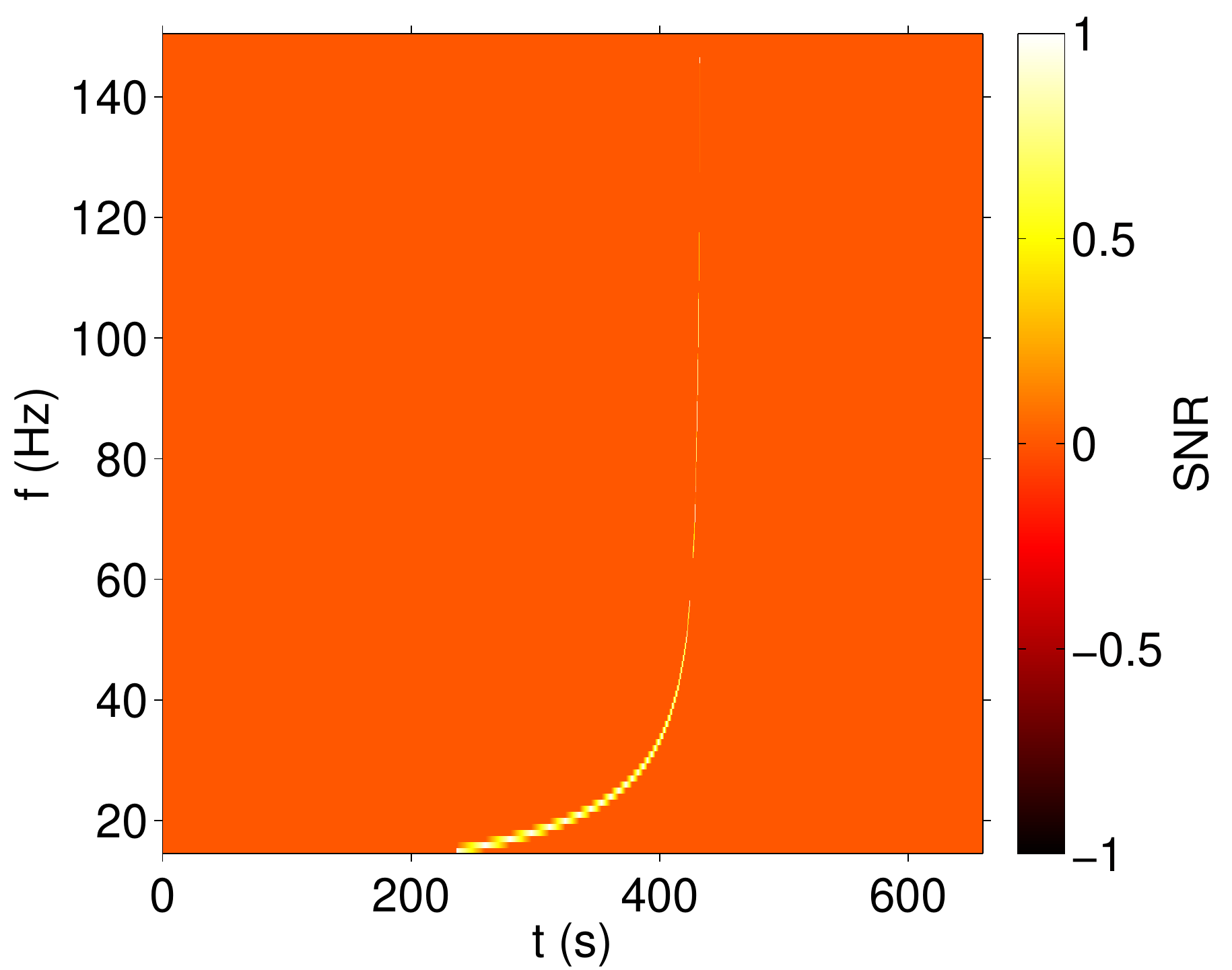}\\
 \includegraphics[width=3.5in]{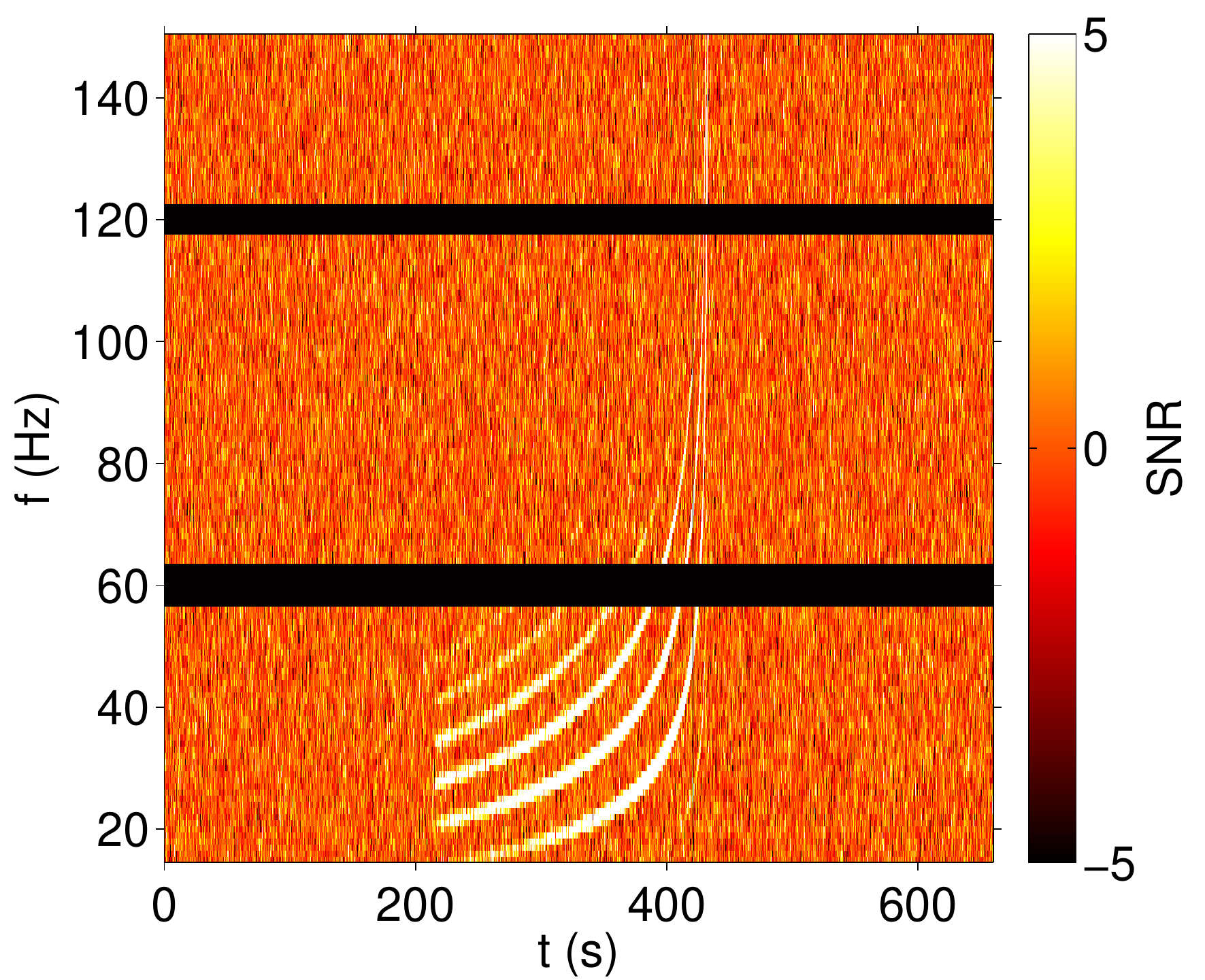}
 \includegraphics[width=3.5in]{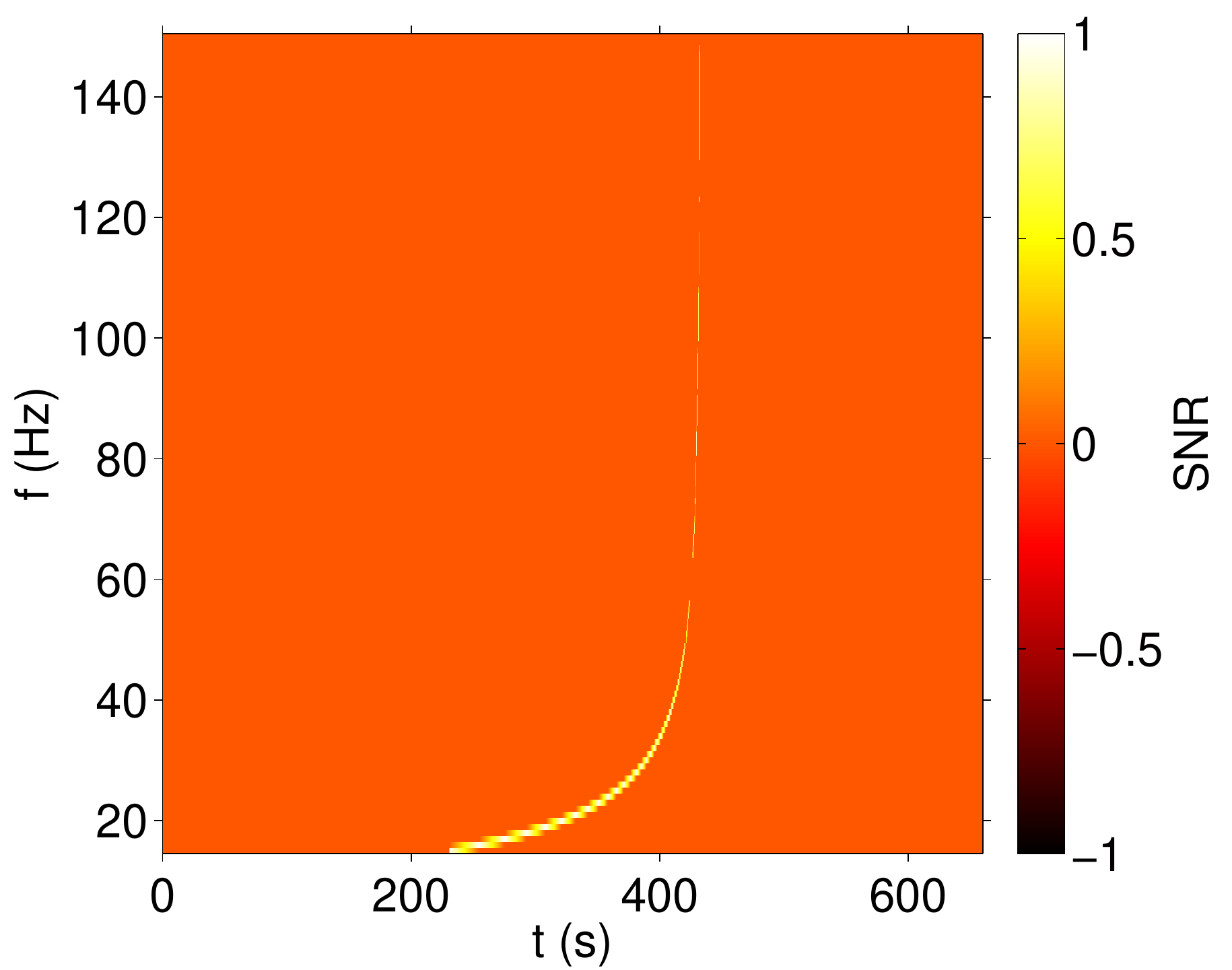}\\ 
 \caption{
   The plot on the top left shows $\rho(t;f)$ for a simulated eccentric ($\epsilon = 0.4$) BNS signal injected on top of Monte Carlo detector noise.
   The simulated noise is created for the advanced LIGO Hanford and Livingston Observatories operating at design sensitivity.
   The distance is $\unit[100]{Mpc}$ and each component mass is $1.4 M_\odot$.
   The gravitational-wave signal appears as a faintly-visible track of lighter-than-average pixels.
   The black horizontal lines are frequency notches used to remove instrumental artifacts.
   On the top right is the recovery obtained with seedless clustering.
   The signal is recovered with a $\text{FAP}<1\%$ (please see section~\ref{sec:Results} for details).
   The plot on the bottom left shows $\rho(t;f)$ for the same system but an order of magnitude nearer at $\unit[10]{Mpc}$. The harmonics are visible at this distance.
   The bottom right shows the recovery obtained with seedless clustering.
 }
 \label{fig:EBNS}
\end{figure*}

In situations where the GW is either difficult to accurately model or the parameter space too large to easily create template banks to accurately span the parameter space, a potential alternative is to search for excess power in spectrograms (also called frequency-time $ft$-maps) of GW detector data \cite{X-Pipeline,CoherentWaveBurst,STAMP}. In these searches, the goal is to design pattern recognition algorithms that can identify the presence of GW signals across the parameter space of interest (in our case, low mass, low-to-moderate eccentricity CBCs). 
A strategy that has been shown to be effective in searches for long-lived transients is known as ``seedless clustering,'' which integrates the signal power along spectrogram tracks using pre-defined ``templates'' chosen to capture the salient features of a wide class of signal models \cite{PhysRevD.88.083010,stochsky,stochtrack_cbc}. Examples of both nearby and near-detection threshold eccentric BNS signals recovered with a seedless chirping template are shown in Fig.~\ref{fig:EBNS}.

In previous work ~\cite{stochtrack_cbc}, the authors have shown how seedless clustering can be applied to searches for low-mass CBC signals ($M_\text{total}\leq10M_\odot$). A shortcoming of the previous analysis was the use of a circular PN expansion when performing the search for the CBC signals, which is sub-optimal for binaries with even low to moderate eccentricities. In this paper, we show how to apply seedless clustering formalism to efficiently search for eccentric CBC signals.
In section \ref{sec:SeedlessClustering}, we review the basics of seedless clustering and show how the formalism of~\cite{PhysRevD.88.083010,stochsky,stochtrack_cbc} can be tuned to more sensitively detect eccentric CBC signals.
In section \ref{sec:Results}, we determine the sensitivity of seedless clustering algorithms to eccentric CBC signals.
In section \ref{sec:computing}, we describe the computational resources required for realistic searches and compare the algorithms' performance on CPUs and GPUs.
In section \ref{sec:implications}, we discuss the implications of the results to the detectability of O'Leary et al.'s model \cite{ScatteringBlackHoles}.
We present our conclusions and a discussion of topics for further study in section \ref{sec:Conclusion}.

\section{Seedless clustering for chirps}
\label{sec:SeedlessClustering}

Spectrograms proportional to GW strain power are the starting point for most searches for unmodeled GW transients. Pixels are computed by dividing detector strain time series in segments and computing the Fourier transform of the segments. We denote the Fourier transform of strain data from detector $I$ for the segment with a mid-time of $t$ by $\tilde{s}_I(t;f)$. The time and frequency resolution is typically optimized based on the signal morphology. Following~\cite{stochtrack_cbc}, we use $50\%$-overlapping, Hann-windowed segments with duration of $\unit[1]{s}$ and a frequency resolution is $\unit[1]{Hz}$.

Searches for long-duration GW transients construct spectrograms of $ft$-maps of cross-power signal-to-noise ratio using the cross-correlation of two GW strain channels \cite{STAMP}:
\begin{equation}
  \rho(t;f|\hat\Omega) = \text{Re}\left[
    \lambda(t;f)
    e^{2\pi i f \Delta\vec{x}\cdot\hat\Omega/c}
    \tilde{s}_I^*(t;f) \tilde{s}_J(t;f)
    \right] .
\end{equation}
In this expression, $\hat\Omega$ is the GW source direction, $\Delta\vec{x}$ is a vector pointing from detector $I$ to detector $J$, $c$ is the speed of light, and $e^{2\pi i f \Delta\vec{x}\cdot\hat\Omega/c}$ is a direction-dependent phase factor that takes into account the time-delay between the two detectors.
The $\lambda(t;f)$ term is a normalization factor that uses data from neighboring segments to estimate the background at time $t$:
\begin{equation}
    \lambda(t;f) = \frac{1}{\cal N} \sqrt{\frac{2}{P'_I(t;f) P'_J(t;f)}} .
\end{equation}
Here, $P'_I(t;f)$ and $P'_J(t;f)$ are the auto-power spectral densities for detectors $I$ and $J$ respectively in the segments neighboring $t$.
One can find additional details in ~\cite{STAMP,PhysRevD.88.083010,stochsky,stochtrack_cbc}.

Pattern recognition algorithms are used to find signals present in the $ft$-maps. The specific form of the potential GW in the $ft$-map depends on the signal.
Low mass, low-to-moderate eccentricity CBCs appear as chirps of increasing frequency.
For highly eccentric signals, the signal also includes distinct and repeated ``pre-bursts'' that last from minutes to days as the binary evolves from the initial very eccentric phase towards the less eccentric phase \cite{PhysRevD.85.123005}.
It is potentially very challenging to design a search, which includes contributions from these pre-bursts, as their spacing is, in general, poorly constrained.

As described above, seedless clustering identifies clusters of pixels, denoted $\Gamma$, likely to be associated with a GW signal by integrating along tracks of pre-defined templates. We denote the number of pixels in $\Gamma$ by $N$. The total signal-to-noise ratio for $\Gamma$ is then a sum over $\rho(t;f|\hat\Omega)$:
\begin{equation}\label{eq:sum}
  \text{SNR}_\text{tot} \equiv
  \frac{1}{N^{1/2}}
  \sum_{\left\{t;f\right\}\in\Gamma} \rho(t;f|\hat\Omega) ,
\end{equation}

While seed-based algorithms connect statistically significant seed pixels to form clusters~\cite{PhysRevD.88.083010,burstegard}, seedless clustering uses banks of parametrized frequency-time tracks. Because of this, calculations for many templates can be performed in parallel, which facilitates rapid calculations on multi-core devices such as graphical processor units (GPUs). 
Because these banks do not use phase and GW waveform amplitude, searches utilizing them are less sensitive than traditional matched filtering searches. On the other hand, for this same reason, they can be more robust when searching for GWs that to not fit the signal model exactly.

There are a number of seedless clustering parameterizations at this point in the literature~\cite{bezier,stochsky,stochtrack_cbc}.
One of the most robust is a template bank of randomly generated B\'ezier curves~\cite{bezier}, which have been shown to be sensitive to a number of long-lived narrowband GW signals~\cite{stochsky}. In the case of circular CBC signals, the most appropriate choice is a PN expansion of the form:

\begin{equation}\label{eq:f_of_t}
  f(t) = \frac{1}{2 \pi} \frac{c^{3}}{4 G M_\text{total}} \sum_{k=0}^7 p_k \tau^{-(3+k)/8} ,
\end{equation}
where
\begin{equation}
  \tau = \frac{\eta c^{3} (t_c -t)}{5 G M} .
\end{equation}
Here, $M_\text{total}$ is the total mass of the binary, $G$ is the gravitational constant, and the expansion coefficients $p_k$ can be found in \cite{PN}. In \cite{stochtrack_cbc}, these were shown to fit the frequency evolution of the circular CBCs very well. On the other hand, the fits for eccentric CBCs were less precise. This motivates the derivation presented below.

To derive an expression for eccentric CBC signals, we use a model from \cite{escidoc:795602}, which uses a PN model that was calibrated by comparison to a numerical relativity (NR) simulation of an eccentric, equal-mass binary black hole. The derivation can be found in Appendix~\ref{sec:EccentricDerivation} and the frequency evolution in equation~\ref{eq:foft}. One characteristic of seedless clustering is the use of a single track across the $ft$-map. This is suboptimal for eccentric signals as some power is present in harmonics of the orbital frequency of the binary, as can be seen in Fig.~\ref{fig:EBNS}. Yunes et al. \cite{PhysRevD.80.084001} show that for small eccentricities, the power is dominated by components oscillating at once, twice and three times the orbital frequency. In the limit $\epsilon << 1$, the dominant term is the second harmonic. It is consistent with the assumptions above then that we search for a single, dominant harmonic with our algorithm.

In~\cite{stochtrack_cbc}, circularized CBC waveforms are parameterized by  two numbers: the coalescence time and the chirp mass.
(In \cite{stochtrack_cbc}, we showed how approximating the individual component masses as equal led to excellent track fits, allowing for a significant reduction in the number of templates required to span the space.)
The inclusion of eccentricity expands the chirp parameter space by an additional dimension.  
In the analysis below, we conservatively use a minimum componenent mass of $1 M_\odot$.

By searching over 40 different time-delays, corresponding to 40 different sky rings, a computationally efficient all-sky search can be performed. This was demonstrated in \cite{stochtrack_cbc} to be sufficient to recover CBC signals in arbitrary directions. The assumption of low-to-moderate eccentricities here leads to a small increase in the number of templates required to span the space of interest. At eccentricities of 0.5 or above, contributions from the neglected terms in the derivation become significant at the 10\% level. Also, the assumption that most of the power is in a single harmonic begins to break down.
Therefore, we search from $0\leq\epsilon\leq0.5$. This leads to an increase in templates by a factor of 6 (using steps of 0.1 in eccentricity).
The sensitivity does not improve appreciably with a higher resolution scan over eccentricity bins.

In order to justify expanding the parameter space, which not only requires more templates but also increases the noise background distribution (requiring higher signal-to-noise ratio to make detections), the fit of the templates must improve. We show below that there is a portion of the parameter space where the eccentric templates have more overlap with the signals and consequently capture more signal-to-noise ratio (SNR), and thereby extend the sensitive distance of the search. A simple metric for determining the efficacy of the fits is the overlap
\begin{equation}
  O(s,h) = \frac{(s|h)}{(s|s)} ,
\end{equation}
where
\begin{equation}
(x|y) = \frac{\sum_{i=1}^{N_x} \rho_x}{\sum_{i=1}^{N_y} \rho_y}
\end{equation}
and $N_x$ and $N_y$ are the number of pixels in the $x$ and $y$ tracks respectively. The overlap corresponds to the sum over the true track $y$ by the template track $x$, as in Eq.~\ref{eq:sum}. 
$O(s,h)=1$ corresponds to perfect overlap whereas $O(s,h)=0$ corresponds to zero overlap.
It is important to note that this definition of overlap is only {\em analogous} to the standard definition for matched filter searches---see, e.g., \cite{PhysRevD.81.024007,PhysRevD.86.084017}---as this definition is designed for spectrographic excess power searches.
The fitting factor, $FF$, gives the loss in SNR due to non-optimal templates. $FF$ is computed by maximizing the overlap function over the template bank

\begin{equation}
  FF(s,h) = \textrm{max} \frac{(s|h)}{(s|s)} .
\end{equation}
Like our expression for $O(s,h)$, $FF(s,h)$ is analogous (but not directly comparable to) definitions from matched filtering; see, e.g., \cite{PhysRevD.81.024007,PhysRevD.86.084017}.
$FF = 1$ means that the fit in templates is perfect, while $FF = 0$ means that there is no overlap. 

\section{Sensitivity study}
The design of the sensitivity study is as follows. We use Monte Carlo Gaussian noise consistent with the design sensitivity of advanced LIGO. We perform an untriggered search over a week of data. Following \cite{stochtrack_cbc}, we create $\unit[660]{s}$ non-overlapping spectrograms. To estimate background, we perform 100 time-slides of a week of data.
In each $ft$-map, we search for a chirp signal using circular templates and using eccentric templates.

We begin our study by determining our background. Using many noise realizations, we estimate the the distribution of $\text{SNR}_\text{tot}$ for the two search variations corresponding to circular and eccentric templates.
We use these noise distributions to determine the value of $\text{SNR}_\text{tot}$ (for each search variation). This corresponds to a false alarm probability (FAP) of $1\%$ for the untriggered searches. 

We then determine the distance at which the signals can be detected with $\text{SNR}_\text{tot}$ sufficient for a $\text{FAP}<1\%$. To do so, we inject GW signals into many noise realizations, with optimal sky location and an optimal source orientation, and recover them with the two search variations. We define the ``sensitive distance'' as the distance at which $50\%$ of the signals are recovered with $\text{FAP}<1\%$ for each pipeline. Following~\cite{stochtrack_cbc}, we use $15$ CBC waveforms with component masses ranging from $1.4$--$5 M_\odot$ and eccentricites ranging from $0-0.6$.
The circular waveforms are generated using a SpinTaylorT4 approximation, while the eccentric waveforms are generated using {\tt CBWaves}, which uses all the contributions that have been worked out for generic eccentric orbits up to 2PN order \cite{CBWaves}.
We provide the parameters for each waveform are give in Table~\ref{tab:Results}.

\section{Results}\label{sec:Results}

\begin{table*}[t]
\begin{tabular}{|c|c|c|c|c|c|c|c|c|c|c|}
\hline
Waveform & $m_1$ & $m_2$ & $\epsilon$ & $t_\text{dur}$ (s) & $FF_\text{Circular}$ & $FF_\text{Eccentric}$ & $D_\text{Circular}$ & $D_\text{Eccentric}$ \\\hline\hline
BNS 1 & 1.4 & 1.4 & 0 & 170 & 0.95 & 0.95 & 160 & 160 \\\hline
NSBH 1 & 3.0 & 1.4 & 0 & 96 & 0.9 & 0.9 & 270 & 270 \\\hline
BBH 1 & 3.0 & 3.0 & 0 & 54 & 0.95 & 0.95 & 420 & 390 \\\hline
BBH 2 & 5.0 & 5.0 & 0 & 42 & 0.75 & 0.75 & 620 & 560 \\\hline
EBNS 1 & 1.4 & 1.4 & 0.2 & 120 & 0.85 & 0.85 & 150 & 160\\\hline
EBNS 2 & 1.4 & 1.4 & 0.4 & 224 & 0.65 & 0.85 & 150 & 160\\\hline
ENSBH 1 & 3.0 & 1.4 & 0.2 & 69 & 0.95 & 0.85 & 240 & 240\\\hline
ENSBH 2 & 3.0 & 1.4 & 0.4 & 127 & 0.6 & 0.9 & 220 & 240\\\hline
ENSBH 3 & 3.0 & 1.4 & 0.6 & 237 & 0.65 & 0.75 & 220 & 240\\\hline
EBBH 1 & 3.0 & 3.0 & 0.2 & 40 & 0.3 & 0.75 & 240 & 360\\\hline
EBBH 2 & 3.0 & 3.0 & 0.4 & 70 & 0.25 & 0.7 & 180 & 290\\\hline
EBBH 3 & 3.0 & 3.0 & 0.6 & 128 & 0.3 & 0.6 & 200 & 350\\\hline
EBBH 4 & 5.0 & 5.0 & 0.2 & 14 & 0.45 & 0.85 & 350 & 420\\\hline
EBBH 5 & 5.0 & 5.0 & 0.4 & 26 & 0.3 & 0.65 & 290 & 390\\\hline
EBBH 6 & 5.0 & 5.0 & 0.6 & 51 & 0.4 & 0.55 & 240 & 390\\\hline
\end{tabular}
\caption{
  Sensitive distances for CBCs with different parameters (assuming optimal sky location and source orientation) given the design sensitivity of advanced LIGO~\cite{obs_scen}.
  There is one row per waveform, and we use the following abbreviations: BNS=``binary neutron star,'' NSBH=``neutron star black hole binary,'' BBH=``binary black hole.''
  Eccentric aveforms begin with an ``E.''
  The component masses in units of $M_\odot$ are given in columns $m_1$  and $m_2$.
  The next columns list the eccentricity $\epsilon$ and the waveform duration in seconds.
  The next two columns list the fitting factors (FF) for both chirp-like templates and eccentric templates.
  The final two columns list the ($\text{FAP}=1\%$, $\text{FDP}=50\%$) detection distance (in Mpc) for circular templates and eccentric templates.
}
\label{tab:Results}
\end{table*}

We summarize the results of our sensitivity study in Table~\ref{tab:Results}.
  First, we evaluate the improvement in sensitivity gained by using eccentric templates.
  Then, for the sake of completeness, we consider the (small) {\em loss} in sensitivity for circular signals due to the expanded search space.

\subsection{Recovery of eccentric signals}
\begin{figure}[t]
	\includegraphics[width=3.5in]{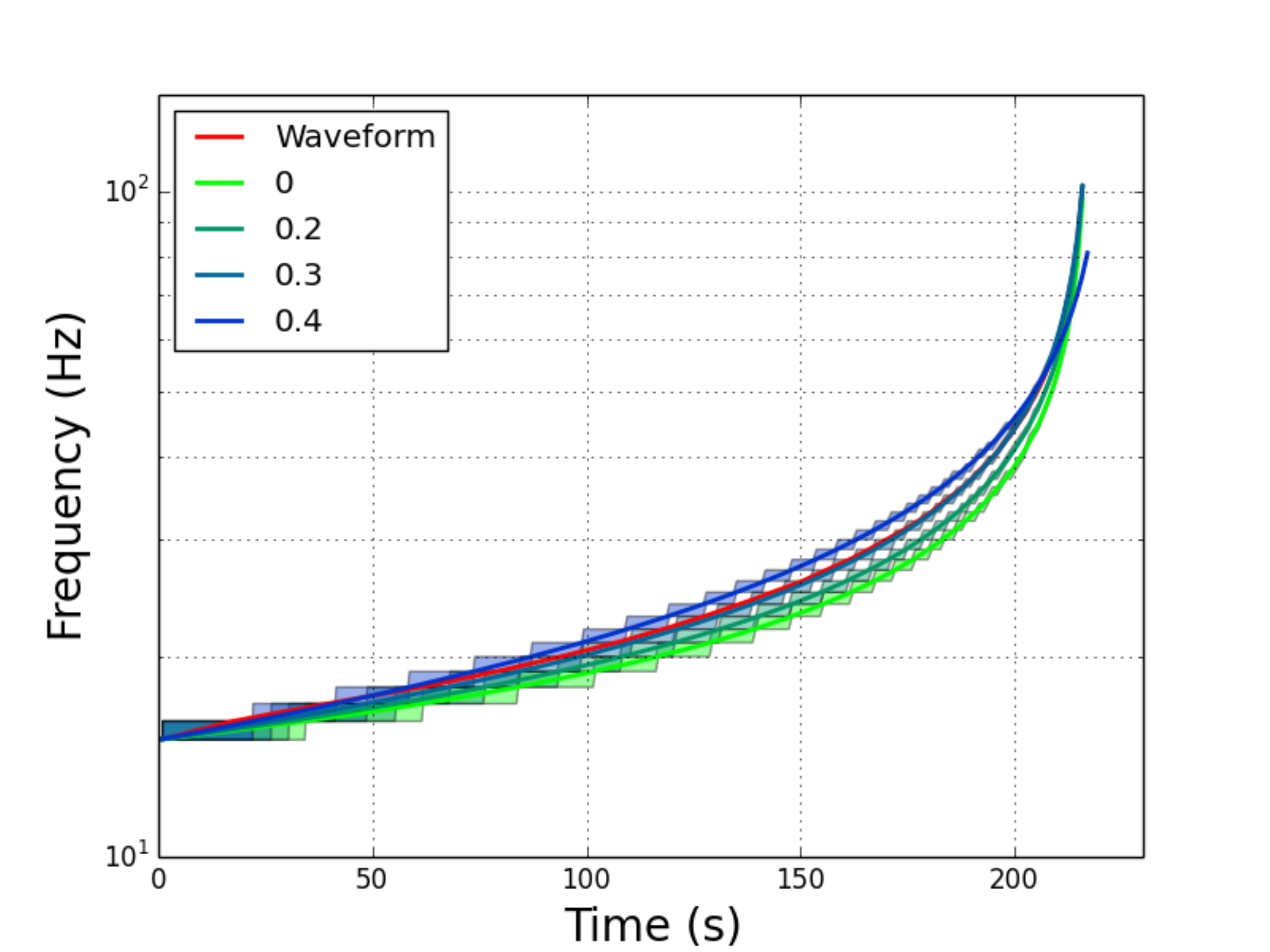}
	\caption{This plot shows a BNS waveform (red) with $\epsilon=0.4$ and $m_1=m_2=\unit[1.4]{M_\odot}$ compared to eccentric templates using Equation~\ref{eq:foft} with eccentricities ranging from $\epsilon=0$ (light green) through $\epsilon = 0.4$ (blue). The shaded regions represent a nominal \unit[1]{Hz} band around the track. Ideally, the shaded region for a track would overlap with the red waveform. While the best fit (which happens for $\epsilon=0.3$) is not associated with the correct eccentricity, it is nonetheless a reasonable fit to the waveform that is better than could be obtained using only a circular parameterization ($\epsilon = 0$). This exhibits the superiority of the shape of the eccentric parameterization in fitting eccentric waveforms.
	}
	\label{fig:fit}
\end{figure}

We begin with an analysis of the fitting factors.
The fitting factors for the eccentric templates range from $0.55-1$, while the fitting factors for the circular templates range from $0.25-1$.
In general, the fitting factors for the circular and eccentric templates are similar for the non-eccentric cases considered here. This is to be expected as the eccentric templates should converge to the circular ones in the limit of small eccentricities.
The performance at low eccentricities for the low-mass systems shows a slightly better fitting factor for the circular templates.
As eccentricity and mass increase, the fitting factors for eccentric systems become significantly higher than that for the circular templates. For the highest mass and most eccentric system considered here, the fitting factors for the eccentric templates can be more than a factor of two higher than for the circular templates. This means that the signal-to-noise ratio recovered for these tracks will be more than a factor of two higher for the eccentric bank. 
Figure~\ref{fig:fit} shows an example of the improvement of eccentric waveform fits from using the eccentric templates.
The main reason that the fitting factors become relatively poor as mass increases is the breakdown of the assumption that the signals being searched for are narrowband. For systems with non-zero eccentricity, there is a broad spectrum of frequencies rather than an identifiable $f(t)$.

The ratio of detection distance for eccentric templates to circular templates ranges from $100$--$175\%$ with a mean of $130\%$. 
This corresponds to an average ratio of sensitive volumes, which corresponds to distance cubed, is $250\%$. 
This is comparable to the gain in sensitivity when going from B\'ezier curves to circular templates \cite{stochtrack_cbc}.
The sensitivity distances for eccentric waveforms decreases as eccentricity increases. The fall-off is significantly slower for the eccentric templates. The breakdown of the circular binary approximation becomes more pronounced at higher eccentricities.
The increase in sensitivity when going from an all-sky search to a triggered search is between $10$--$20$\% for both circular and eccentric templates.


\subsection{Recovery of circular signals}
We now turn our attention to the sensitivity distances.
The trends in the sensitivity distances are similar to those seen in the fitting factors. In general, the sensitivity to the circular waveforms for the eccentric templates is slightly worse than the circular waveforms. This is predominantly due to the increased background distribution from the expansion of the parameter space.
The ratio of detection distance for eccentric templates / circular templates ranges from $90$--$100\%$ with a mean of $95\%$. The average ratio of sensitive volumes is $88\%$.

\section{Computational requirements}
\label{sec:computing}

To estimate the computational cost of an all-sky seedless clustering search (with eccentric chirp-like templates), we performed a benchmark study using a Kepler GK104s GPU and an 8-core Intel Xeon E5-4650 CPU.
We allotted $\unit[8]{g}$ of memory to each job.
The GPU was able to analyze $\unit[660]{s}$ of data in $\unit[106]{s}$, corresponding to a duty cycle of $\approx16\%$. This is about a factor of two slower than the circular-template search.
The CPU duty cycle was comparable using all eight cores, and the job-by-job variability in run time is greater than the difference between using a GPU or an 8-core CPU on average.

For background estimation at the level of $\text{FAP}=1\%$, which corresponds to performing 100 time-slides, it follows that a continuously running seedless clustering search with chirp-like templates can be carried out with just 32 continuously-running GPUs (or 8-core CPUs).
Here, we have taken into account an additional factor of two needed to implement overlapping spectrograms to ensure that signals do not fall on the boundary.
(320~GPUs / 8-core CPUs would be required for background estimation at the level of $\text{FAP}=0.1\%$)
In a real science run, the duty cycle from coincident GW detectors is likely to be $\approx50\%$. This would make these computing requirements conservative by a factor of two.

\section{Astrophysical implications}
\label{sec:implications}

O'Leary et al.~present a model where CBCs with high eccentricities, corresponding to eccentricities near to 1, are produced by the scattering of stellar mass black holes in galactic cores which contain a super-massive black hole \cite{ScatteringBlackHoles}. We discuss here the potential for detecting such systems given the rates presented in the paper. Table I of \cite{ScatteringBlackHoles} presents the Merger Rate per Milky Way Equivalent Galaxy (MWEG) for the models considered. We can straightforwardly convert from the rates in this paper to what we expect in the advanced detector era. The most optimistic scenario, given by model F-1, predicts $1.5 \times 10^{-2}\,\text{MWEG}^{-1}\,\text{Myr}^{-1}$. The median scenario, given by the median of the models considered, predicts $3.3 \times 10^{-4}\,\text{MWEG}^{-1} \text{Myr}^{-1}$. The pessimistic scenario, given by model A$\beta$3, predicts $2.0 \times 10^{-4}\,\text{MWEG}^{-1} \text{Myr}^{-1}$. We now find the rates of CBCs given by Table II of \cite{rates}. The low, realistic, and high rates for BBHs are $0.01\,\text{MWEG}^{-1} \text{Myr}^{-1}$, $0.4\,\text{MWEG}^{-1} \text{Myr}^{-1}$, and $30\,\text{MWEG}^{-1} \text{Myr}^{-1}$ respectively. The matched filter detection rates of CBCs for advanced LIGO are given by Table IV of \cite{rates}. The low, realistic, and high detection rates for BBHs are 0.4, 20, and 1000 respectively. 

\begin{table}[t]
\begin{tabular}{|c|c|c|c|}
\hline
Algorithm & Low Rate & Realistic rate & High rate\\\hline\hline
Matched Filtering & 0.01 & 0.02 & 0.66 \\\hline
Seedless Clustering & 0.001 & 0.002 & 0.06 \\\hline
\end{tabular}
\caption{
  Potential detection rates of eccentric compact binary coalescences for both matched filtering and seedless clustering. The rates combine results for eccentric binaries given by O'Leary et al. \cite{ScatteringBlackHoles}, as well as aLIGO detection rates. The matched filtering line assume a dedicated eccentric binary matched filtering pipeline, which does not currently exist. The seedless clustering algorithm is the one presented in this paper. Please see the text of section~\ref{sec:implications} for further details.
}
\label{tab:Rates}
\end{table}

Using these estimates, we can compute the expected rates of eccentric detections by a matched filtering pipeline using the ratio of the O'Leary et al. and aLIGO detection rates, multiplied by the BBH detection rate. In the case of an electromagnetic trigger, matched filter pipelines improve by about 33\% in volume. These are: 0.01, 0.02, and 0.66 for the low, realistic, and high detection rates for BBHs. We can convert between matched filter and seedless clustering detection rates by dividing through by 8 (as the distances differ by about a factor of 2) \cite{stochtrack_cbc}. These are given by 0.001, 0.002, and 0.06 for the low, realistic, and high detection rates for BBHs. Table~\ref{tab:Rates} summarizes these results. We note here that the matched filtering results stated here would require a dedicated eccentric binary matched filtering pipeline, which does not currently exist. A matched filtering pipeline using circular templates would have rates similar in order of magnitude to that of seedless clustering. We describe in section~\ref{sec:Results} where seedless clustering is most competitive. The relatively low detection rates are due to the significantly fewer eccentric binaries expected relative to circular binaries, at least in the O'Leary et al. model.

\section{Discussion}\label{sec:Conclusion}
We have described an analytic expression for the frequency evolution of low-mass, low-to-moderate eccentricity waveforms.
We showed how an implementation of this evolution for seedless clustering, optimized for eccentric compact binary coalescences, can improve searches for eccentric signals significantly.
We find that the eccentric search can expand the sensitive volume by as much as a factor of $3\times$ depending on the waveform (a factor of 1.4 on average) compared to a comparable circular search.

In the event that circular template banks are used by matched filtering to search for eccentric signals \cite{PhysRevD.87.082004}, there will be a non-negligable loss in sensitivity for these searches. For BNS systems with eccentricities of 0.2 and 0.4, Huerta and Brown estimate signal-to-noise ratio loss factors of about 0.5 and 0.2 respectively \cite{PhysRevD.87.127501}. This would bring the matched filtering sensitivity distances of these signals to $\unit[225]{Mpc}$ and $\unit[90]{Mpc}$, compared to $\unit[180]{Mpc}$ and $\unit[180]{Mpc}$ for seedless clustering; therefore seedless clustering with eccentric template banks may provide further opportunities for observing these types of signals.

In the future, we intend to explore the possibility that including amplitude information in the track recoveries improves the detection sensitivities.
This could be beneficial for compact binaries because the amplitude information is known.
The idea would be to weight spectrogram pixels predicted to contain a higher amplitude of SNR more strongly than those predicted to contain less.
This would weight the relatively low-SNR contribution at low frequency less and thereby decrease the background SNR distribution.
Also, it will be useful to carry out a systematic comparison of seedless clustering with matched filtering pipelines using non-Gaussian noise. 
Finally, we intend to use this algorithm on future data from advanced LIGO and advanced Virgo.

\section{Acknowledgments}
MC is supported by National Science Foundation Graduate Research Fellowship Program, under NSF grant number DGE 1144152.
ET is a member of the LIGO Laboratory, supported by funding from United States National Science Foundation.
NC and JL's work was supported by NSF grant PHY-1204371.
LIGO was constructed by the California Institute of Technology and Massachusetts Institute of Technology with funding from the National Science Foundation and operates under cooperative agreement PHY-0757058.
We greatly appreciate the helpful comments about the frequency evolution of eccentric compact binary coalescences from Ilya Mandel.
This paper has been assigned document number P1400242.

\begin{appendix}
\section{Eccentric template derivation}
\label{sec:EccentricDerivation}

To derive an expression for eccentric CBC signals, we use the x-model from \cite{escidoc:795602}, which uses a PN model that was calibrated by comparison to a numerical relativity (NR) simulation of an eccentric, equal-mass binary black hole. This model is named after the choice of coordinates used to express the PN equations of motion, namely, the angular velocity of the compact objects, where $x = (M \omega)^{2/3}$.
To derive an analytic solution, we keep the lowest order terms in $x$ for both the $x$ and $\epsilon$ evolution. Because we are fitting the frequency evolution of the inspiral, the fact that the PN calculations tend to slowly converge at late inspiral are less important here. The differential equations are:
\begin{gather}
\dot{x} = \frac{2 \eta}{15 (1 - \epsilon^2)^{7/2}} (96 + 292 \epsilon^2 + 37 \epsilon^4) x^5 + \mathcal{O}(x^6) \\
\dot{\epsilon} = \frac{-\epsilon \eta}{15 (1 - \epsilon^2)^{5/2}} (304 + 121 \epsilon^2) x^4 + \mathcal{O}(x^5) ,
\end{gather}
where $x \equiv (m*\omega)^{2/3}$.
Taking the ratio of these equations, we obtain
\begin{equation}
  \frac{dx}{d\epsilon} = \frac{-2}{\epsilon}\left(\frac{1}{1-\epsilon^2}\right) \left(\frac{96 + 292 \epsilon^2 + 37 \epsilon^4}{304 + 121 \epsilon^2} \right) x .
\end{equation}
We integrate this equation, yielding
\begin{equation}
  x(\epsilon) = C_0 \left[\frac{1-\epsilon^2}{\epsilon^{12/19} (304 + 121 \epsilon^2)^{870/2299}} \right] .
\end{equation}
Plugging this equation back into the original differential equation and expanding to fourth order in epsilon, results in
\begin{equation}
  \epsilon(t) = \left(\frac{B-t}{A \times M}\right)^{19/48} ,
\end{equation}
where
\begin{gather}
  A = \frac{5 \times 31046}{172 \times 2^{2173/2299} \times 19^{1118/229} \eta C_0^4} \\
  C_0 = (2 \pi M f_0)^{2/3} \left[\frac{1-\epsilon_{0}^2}{\epsilon_{0}^{12/19} (304 + 121 \epsilon_{0}^2)^{870/2299}} \right]^{-1} \\\textsl{}
B = A M \epsilon_0^{48/19} .
\end{gather}
Combining these together,
\begin{equation}
  f(t) = \frac{1}{2 \pi M} \left[\frac{C_0 (1-\epsilon^2)}{\epsilon^{12/19} (304 + 121 \epsilon^2)^{870/2299}} \right]^{3/2} ,
	\label{eq:foft}
\end{equation}
where $f_0$ is the initial frequency of the binary, and $\epsilon_{0}$ is the initial eccentricity.

\end{appendix}

\bibliography{stochtrack_ecbc}

\end{document}